# Title: Inverse Doppler Effects in Flute

**Authors:** Xiao P. Zhao*, Shi L. Zhai, Song Liu, Fang L. Shen, Lin L. Li and Chun R. Luo

**Affiliations:**

Smart Materials Laboratory, Department of Applied Physics, Northwestern Polytechnical University, Xi' an 710129, People's Republic of China.

*Corresponding author. E-mail: xpzhao@nwpu.edu.cn

**Abstract**: Here we report the observation of the inverse Doppler effects in a flute. It is experimentally verified that, when there is a relative movement between the source and the observer, the inverse Doppler effect could be detected for all seven pitches of a musical scale produced by a flute. Higher tone is associated with a greater shift in frequency. The effect of the inverse frequency shift may provide new insights into why the flute, with its euphonious tone, has been popular for thousands of years in Asia and Europe.

**One Sentence Summary:** The inverse Doppler effect in a flute could be detected for all seven pitches of a musical scale produced, and higher tone is associated with a greater shift in frequency.

**Main Text:**

The Doppler effect is a fundamental phenomenon in wave propagation. In 1843, Doppler first generalized the change in frequency of a wave for an observer moving relative to its source (*1*). Nowadays, this phenomenon is utilized in many fields, including space technology, traffic management, medical diagnosis, to name a few. In 1968, Veselago (*2*) theoretically predicted that materials with negative refractions are capable of realizing the inverse Doppler effect. Recent progress on metamaterials, especially the invention of photonic crystal, provided experimental validation to Veselago's theory (*3−7*).

Flutes, as one of the earliest known instruments, have existed for 8000 years (*8*), perhaps more early (*9*) because of their simple construction, ease of use, and euphonious tone. Although previous generations of musicians did not know the physical mechanisms of a resonator, they observed the relationships of the length and the opening manner of flute with sound pitch out from it. The materials of flutes have evolved from ancient animal femurs to bamboo or metal in modern days, and their manufacturing process has become fairly sophisticated. Research shows that the tube length and opening manner of a flute correspond to its resonant frequency. However, little attention has been paid to the sound effect caused by the relative movement between a flute and the audience.

In previous research, our group experimentally presented a three-dimensional flute-model meta-molecular structure acoustic metamaterial with a periodic array of perforated hollow steel tubes in an impedance tube system (*10*) and a hollow plastic tube with a side hole in free space (*11*). The refractive index, effective mass density, and effective bulk modulus of this setup were found to be negative in a specific frequency range by retrieving the effective parameters from experimental data. Inspired by the fact that the acoustic metamaterial composed of "flute-like" meta-molecule can realize inverse Doppler effects, we hypothesized that the flute also features the inverse Doppler effect. We experimentally measured the Doppler behavior of a flute using a standard acoustic test method (Supplementary materials), Figure. 1 shows the test results. The frequency detected by a moving observer exhibits inverse Doppler behavior compared with that generated by the flute in different tones. As the observer approaches the phonation position, the received frequency becomes lower; conversely, as the distance between the observer and flutist increases, the received frequency becomes higher. The refractive indices derived from the measured Doppler shifts are negative, as shown in Table. S1. Although differences in sound intensity and frequency may be observed between the blow hole and finger holes, the seven tones obtained from the blow hole and finger holes all exhibit inverse Doppler effects. Calculations show that, as the inverse Doppler effect appears, the refractive index of the flute over different tones becomes negative.

"Flute-like" artificial meta-molecules simultaneously show a negative bulk modulus and mass density because of local resonance and can realize many phenomena, such as negative refractions and inverse Doppler effects. Our experiments demonstrate that all seven tones obtained from the flute exhibit inverse Doppler effects. Here, the frequency shift increases with the frequency change of every tone. Although the positions of a musician and his or her audience are fixed, relative movements occur between the blow hole or the finger holes (sound source) and the ears (receiver) of the audience. These movements include reciprocating motion in the horizontal direction, motion in the vertical direction, front and back motion, and even curvilinear motion. As the musical composition changes, the pose and velocity of movement of a musician frequently changes. It Is known that the sound resolution of human ears can reach 0.1 Hz, and at the frequency range below 1000 Hz, ordinary people can easily distinguish the frequency change of 3 Hz. Our experiments demonstrate that when the moving speed is 0.5 m/s, the frequency change of every tone meets the requirement of distinguishability.

    The body movements of musicians are generally believed to express their emotions and catch the attention of the audience (Movies. S1). The results of our research reveal that musicians may consider the Doppler effect caused by these motions, i.e., the effect of frequency shift between source and receiver. Flutes and many wind instruments, such as the clarinet, oboe, and suona horn, can be used to play dulcet music, likely because the relation between sounds created by these instruments and those received by the audience features inverse Doppler effects. While the Doppler effect was first presented about 200 years ago, the property of inverse Doppler effect has been utilized for thousands of years. Flutes may be the oldest instrument associated with a negative refractive index in sound wave propagation. The flute can be a basic unit for the design of many acoustic metamaterials. The human civilization is miraculous.

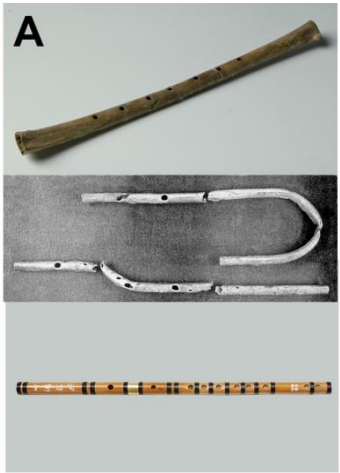
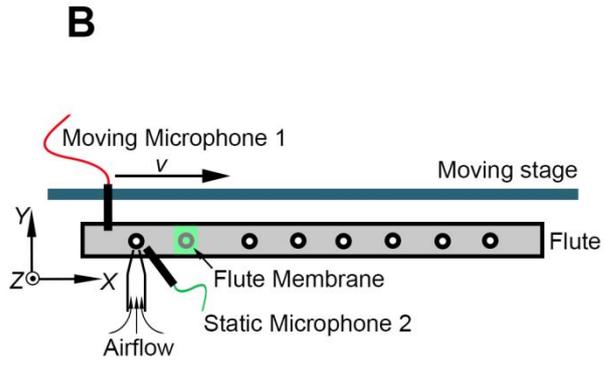
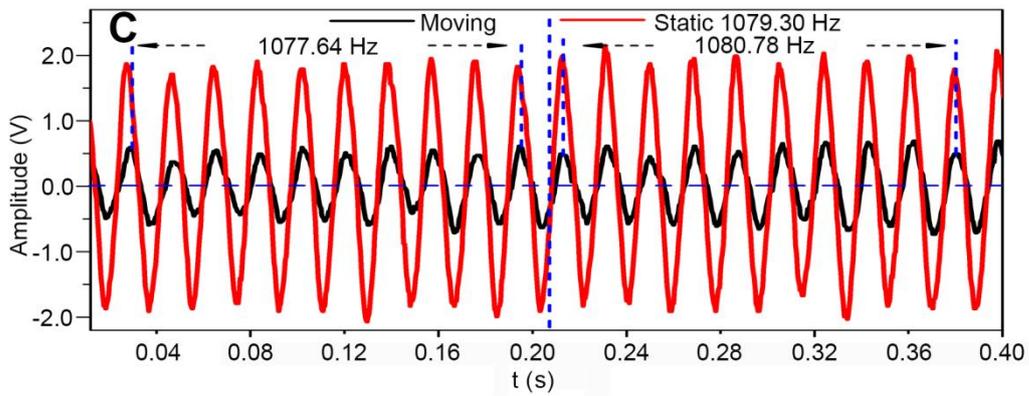
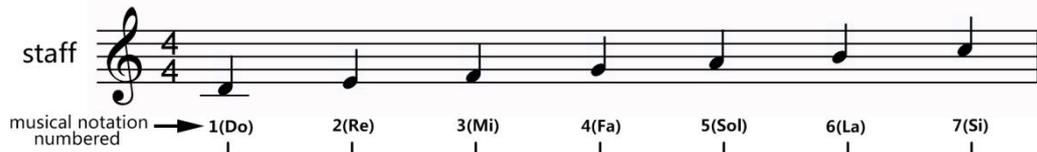
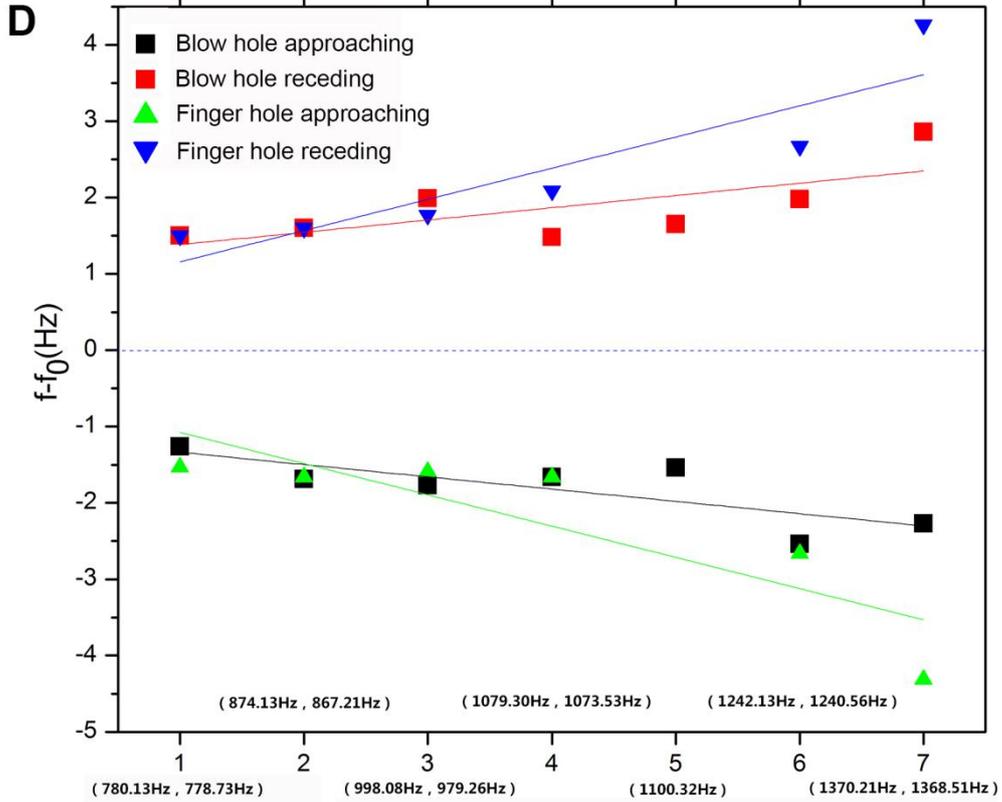

**Fig. 1.** Inverse Doppler effect of the flute. (**A**) Photographs of the ancient and modern flutes. From top to bottom, they are a bone flute excavated in Jiahu China in 1986 and existed for 8000 years (*8*), a silver flute excavated in Sumer in the Southern Mesopotamian valley in present-day Iraq in 1922 and existed for 4500 years (*12*), and a modern flute, respectively. (**B**) Sketch map of the experimental setup for the Doppler shifts of flute. (Supplementary material 4. Measurements of the Doppler effect of the flute) (**C**) Oscillograms detected by the static and moving microphones for the tone 4 at the blow hole. The frequency of source is 1079.3 Hz. When the moving microphone approaches the source, the detected frequency is reduced by 1.51 Hz; as the microphone recedes, the frequency increases by 1.26 Hz. (**D**) Doppler shifts of the flute in different numbered musical notations (i.e. 1-7). The black and red squares refer to the frequency shift results of approaching and receding processes, respectively, which are measured near the blow hole. The blue and cyan triangles indicate the frequency shift results of approaching and receding processes, respectively, which are measured near the first opened finger hole. The solid lines refer to the fitting lines. The positive values mean the detected frequency is larger than source frequency, and the negative values imply the contrary situation.

**Acknowledgments:** We thank J. Zhao for fruitful discussions. This work was supported by the National Natural Science Foundation of China (under Grant